\documentclass[12pt]{iopart}
\usepackage{iopams}
\usepackage{setstack}
\usepackage{graphicx}
\usepackage{caption}
\usepackage{subcaption}
\usepackage{subeqn}
\newcommand{\sgn}{\mathop{\mathrm{sgn}}}
\begin{document}

\title{Fast gates for ion traps by splitting laser pulses}

\author{C D B Bentley$^1$, A R R Carvalho$^{1,2}$, D Kielpinski$^3$ and J J Hope$^1$}

\address{$^1$Department of Quantum Science, Research School of Physics and Engineering, Australian National University, Canberra, Australia}
\address{$^{2}$ARC Centre for Quantum Computation and Communication Technology, The Australian National University, ACT 0200, Australia}
\address{$^3$Centre for Quantum Dynamics, Griffith University, Brisbane, Australia}
\ead{christopher.bentley@anu.edu.au}
\begin{abstract}
We present a fast phase gate scheme that is experimentally achievable and has an operation time more than two orders of magnitude faster than current experimental schemes for low numbers of pulses.  The gate time improves with the number of pulses following an inverse power law.  Unlike implemented schemes which excite precise motional sidebands, thus limiting the gate timescale, our scheme excites multiple motional states using discrete ultra-fast pulses.  We use beam-splitters to divide pulses into smaller components to overcome limitations due to the finite laser pulse repetition rate.  This provides gate times faster than proposed theoretical schemes when we optimise a practical setup.
\end{abstract}


\maketitle

\section{Introduction}

Practical quantum information processing (QIP) requires a number of essential components outlined by DiVincenzo in~\cite{DiV00FP}.  Ion trap schemes fulfil most of these criteria, being a particularly promising QIP platform~\cite{CZ95PRL, GZC05JPB, BlWi08Nat, Haff08PR}.  The qubits are represented by the internal states of the ions, which permit high fidelity initial state preparation and readout~\cite{Myer08PRL, NC00}.  The states of the qubits are manipulated using atom-laser interactions, which allow the implementation of a universal set of gates \cite{DiV95PRA}: single-qubit rotations are straightforward, and non-separable multiple-qubit gates are mediated by the shared motional modes of the ions.  The ion trap system has demonstrated its potential with the implementation of teleportation \cite{Barr04Nat, Rieb04Nat, Olms09Sci}, entanglement of up to 14 ions \cite{Turc98PRL, Sack00Nat, Haff05Nat, Monz11PRL}, quantum algorithms \cite{Guld03Nat, Chia04Nat, Bric05PRA, Chia05Sci} and simulation of quantum systems \cite{Barr11Nat, Isla11NC}.  


One important criterion is the requirement of a high ratio between the system's decoherence time and the timescale for quantum gates.  In ion traps decoherence times of up to 10 minutes \cite{Boll91IEEETIM} have been achieved, while gates have been performed with speeds up to the order of 10$\mu$s \cite{DeMa02PRL, Kirch09NJP}, and fidelity up to 0.993(1) \cite{Benh08NP}.  Most implemented gates \cite{DeMa02PRL, Kirch09NJP, Benh08NP, Monr95PRL, Leib03Nat, S-K03Nat, Ospe11Nat}, such as the Cirac-Zoller (CZ) \cite{CZ95PRL} and M{\o}lmer-S{\o}rensen (MS) \cite{SoMo99PRL} gates, excite specific motional sidebands. This restricts the laser intensity that can be used and, consequently, limits the gate time to be much slower than the trap oscillation period \cite{Zhu06EL}. For longer, more complex operations, faster and higher fidelity gates are required.

Fast gate schemes have been proposed where all of the phonon modes are excited \cite{Zhu06EL, Duan04PRLa, GZC03a, GZC05a}, allowing for higher laser power and faster interactions, but complicating the disentanglement of motional modes and internal states.  More sophisticated control is required, using amplitude shaped pulses or carefully timed sequences of pulses. Producing the necessary continuous, shaped pulses is a significant experimental challenge and such schemes are yet to be implemented. Despite being closer to implementation, discrete pulse schemes also face certain limitations. Lasers practical for quantum information processing have repetition rates up to around 300 MHz \cite{Kiel11IQEC} and the pulses are fixed in time according to the repetition rate of the ultra-fast laser. This is a restrictive, or sometimes prohibitive, limitation for pulsed schemes.

In this paper we outline a speed-up of quantum gates for ion traps through pulse sequences that are readily realised experimentally. We consider a pulse splitting technique using unequal path-length interferometers to increase the rate of pulses incident on the ions, for a higher effective repetition rate.  The pulse splitting process also provides certain timing freedoms for the pulse components. These two features overcome the hitherto limiting factors for pulsed fast gate schemes.

The paper is organised as follows: we first consider the fast gate scheme in \cite{GZC03a}, presented by Garc\'{i}a-Ripoll, Zoller and Cirac (GZC scheme), which addresses two ions and is arbitrarily fast for infinitesimal pulse lengths and separations.  The scheme and the limitations in its implementation are presented in section \ref{Gmodel}.  In section \ref{SSplitting}, we introduce the pulse splitting technique and show how it can produce the pulse timings required to experimentally implement the GZC scheme.  In section \ref{SOptim}, we optimise the pulse splitting technique and propose a faster gate scheme than existing theoretical schemes for typical repetition rates, with a simpler experimental setup.  The physical limitations of our scheme are outlined in section \ref{Gerror}.

\section{Garc\'{i}a-Ripoll, Zoller and Cirac (GZC) scheme} \label{Gmodel}

We review the proposed GZC fast gate scheme using discrete pulses given in \cite{GZC03a}.  This scheme offers arbitrarily fast gates given perfect laser control, it does not require ground state cooling and works outside the Lamb-Dicke regime \cite{Lee05JOB}.  The gates are presented in the form of a control problem, where two conditions are imposed:

\begin{description}
\item[Phase condition] \hfill \\ A two-qubit phase gate must be completed, which requires the relationship between the initial and final qubit states to be described by the gate unitary
\begin{eqnarray}
U_I= e^{i \Theta \sigma_1^z \sigma_2^z}. \label{eqidu}
\end{eqnarray}
$\Theta$ is a two-qubit state dependent relative phase, and we consider the case $\Theta = \pi/4$.

\item[Motional mode condition] \hfill \\ The gate should be independent of the motional state, such that ground state cooling is not required.  For a two qubit system, there are two motional modes: the centre of mass and stretch modes.  The initial states of both modes (with free evolution) should be restored at the end of the gate.
\end{description}

The gate is constructed from discrete pulses in such a way that these conditions are satisfied.  We now consider the system evolution to quantify the conditions.

\subsection{Gate details}

The Hamiltonian for two ions interacting with resonant light is
\begin{eqnarray}
H_I = \frac{\Omega(t)}{2}[\sigma_1^+ e^{ikx_1} + \sigma_2^+ e^{ikx_2} + H.c.],
\end{eqnarray}
where $\Omega(t)$ is the Rabi frequency, $x_1$ and $x_2$ are the positions of the two ions, and $\sigma_1^+$ and $\sigma_2^+$ are the raising operators for the internal states of the ions.

The GZC scheme uses $N$ pairs of counter-propagating $\pi$ pulses that address both ions and are on resonance with the internal transition frequency.  Each pulse pair, with negligible delay between pulses, maintains the internal states of the ions while imparting a $2 \hbar k$ momentum kick \cite{Kaza74ZETF}.  The combined evolution operators for free evolution and pulse pairs are:
\begin{subeqnarray} \label{unitcr}
\mathcal{U}_c = \Pi_{k=1}^N e^{-i 2 z_k \eta_c (\sigma_1^z + \sigma_2^z) (a_c + a_c^\dagger)} e^{-i \nu_c \delta t_k a_c^\dagger a_c} \\
\mathcal{U}_r = \Pi_{k=1}^N e^{-i z_k \eta_r (\sigma_1^z - \sigma_2^z)(a_r + a_r^\dagger)} e^{-i \nu_r \delta t_k a_r^\dagger a_r}.
\end{subeqnarray}
The subscripts $c$ and $r$ refer to the centre of mass and stretch (or relative motion) modes respectively, such that $\mathcal{U}_c$ is the unitary operator for the centre of mass mode, and $a_r$ is the stretch mode annihilation operator.  The Lamb-Dicke parameter $\eta$ determines $\eta_c = \eta/\sqrt{2}$ and $\eta_r = \eta \sqrt[4]{4/3}$, while the trap frequency $\nu = \nu_c$, and $\nu_r = \sqrt{3}\nu$.  $|z_k|$ is the number of simultaneous pairs of pulses incident on the ions at a time $\delta t_k$ after the previous group of $|z_{k-1}|$ pulse pairs, where $\delta t_1=0$.  The direction of the first pulse for each pair in the $k^{\mathrm{th}}$ group of pairs is determined by $\sgn{(z_k)}$.

Note that the components of the unitaries correspond to displacement operators for the momentum kicks, interspersed with rotation operators for the free evolution.  For an initial coherent state, the motion of the centre of the coherent state can be described by a classical trajectory in (x,p)-phase space for each mode.  If this phase space trajectory is closed, then the motional state is restored and the motional mode condition is fulfilled.

When phase space is rotated at the relevant mode frequency to cancel the natural rotation of a coherent state, a closed trajectory maps out an area corresponding to a geometric phase factor.  This phase factor determines the conditional phase $\Theta$, which we require to be $\pi/4$ to satisfy the phase condition (equation (\ref{eqidu})).  Up to a global phase factor, this means that the relative phase shift between the computational excited and ground states ($|00\rangle$, $|11\rangle$) and the partly excited states ($|01\rangle$, $|10\rangle$) is $\pi/2$.  The unitaries in equation (\ref{unitcr}) show that the excited and ground states are only displaced in phase space for the centre of mass mode, while the partly excited states are only displaced in stretch mode phase space.  We thus require that the relative difference between the geometric phase factors for each mode is given by $\pi/2$.

As in \cite{GZC03a} we take the product of the consecutive displacement and rotation operators for coherent states, or superpositions thereof, from equation (\ref{unitcr}).  This gives the quantified control conditions:
\begin{subeqnarray}\label{eq:cond}
\Theta &= 4\eta^2 \sum^N_{m=2} \sum^{m-1}_{k=1} z_m z_k \left[\sin(\nu \delta t_{mk}) - \frac{\sin(\sqrt{3}\nu\delta t_{mk})}{\sqrt{3}}\right] = \frac{\pi}{4} \label{Phase} \\
C_c &= \sum^N_{k=1} z_k e^{-i\nu t_k} = 0 \label{Cc}\\
C_r &= \sum^N_{k=1} z_k e^{-i \sqrt{3} \nu t_k} = 0, \label{Cr}
\end{subeqnarray}
where the time of incidence of the $k^{\mathrm{th}}$ group of pairs is $t_k$, and $\delta t_{mk} = t_m - t_k > 0$.  Equation (\ref{Phase}) represents the difference between geometric phase factors as discussed.  $C_c$ and $C_r$ are a measure of the distance between the initial and final trajectory points for the centre of mass and stretch modes respectively, while equations (\ref{Cc}) and (\ref{Cr}) ensure that the trajectories are closed.

The solution space grows exponentially with the number of pulses, thus the schemes satisfying the condition equations presented in \cite{GZC03a} include simple symmetries.  We refer to the GZC gate scheme as the solution to the condition equations (\ref{eq:cond}) presented in \cite{GZC03a} which offers arbitrarily fast gate times given arbitrarily fast laser pulses on demand.  The scheme is characterised by instantaneous groups of pulse pairs $\underbar{z}$ sent at times $\underbar{t}$, interspersed with free evolution: 
\begin{subeqnarray}
\underbar{z} = (-2n,3n,-2n,2n,-3n,2n) \\
\underbar{t} = (-\tau_1,-\tau_2,-\tau_3,\tau_3,\tau_2,\tau_1).
\end{subeqnarray}

As above, at time $t_k \in \underbar{t}$, $z_k \in \underbar{z}$ determines the number of simultaneous counter-propagating pulse pairs $|z_k|$ that are incident on the ions, as well as the direction $\sgn (z_k)$ of the first pulse in each pair.  Every pulse pair in a group has the same first pulse direction.  Note that more than one pulse pair is permitted at a given time $t_k$, and the pulse pairs themselves are considered to take negligible time.  To obtain arbitrarily fast gate times, the number of pulses in the scheme, $N = 14n$, is scaled towards infinity.

\begin{figure}[h!]
 \begin{subfigure}[b]{0.5\textwidth}
     \centerline{\includegraphics[width=\textwidth]{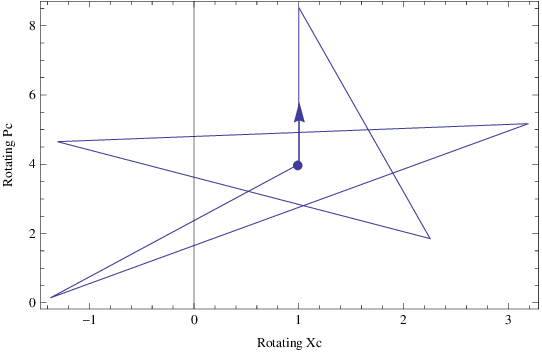}}
   \caption{n=4}
   \label{GZCTraj4}
 \end{subfigure}
 ~
 \begin{subfigure}[b]{0.5\textwidth}
      \centerline{\includegraphics[width=\textwidth]{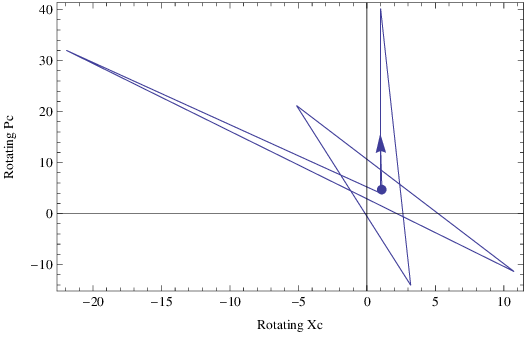}}
   \caption{n=32}
   \label{GZCTraj32}
 \end{subfigure}
   \caption{Phase space trajectories for the GZC gate, where $Xc$ and $Pc$ are the dimensionless position and momentum for the centre of mass mode.  Here phase space is rotating with frequency $\nu$.  Increasing numbers of pulses, $n$, reduce the time evolution, seen through smaller angles in the second plot.  The larger kicks are also apparent, maintaining the required geometric phase.  The initial state and trajectory direction are shown.}
   \label{fig:GZCTrajs}
\end{figure}

Figure \ref{fig:GZCTrajs} shows the rotating phase space trajectory plots for the centre of mass mode with increasing $n$.  Phase space trajectories identify the behaviour of a given solution scheme, as well as providing a check on the scheme validity since the closure of trajectories and geometric phase are apparent.  Note that as the number of pulses, or kicks, is scaled towards infinity, the angles (evolution times between kicks) shrink as the kick sizes increase, preserving the area and maintaining the closed trajectory.  The scheme's star shape aids in this effective scaling with $n$.


\subsection{GZC limitations}

Gate times are restricted by dissipation in the system, due to coupling to the environment, as well as control errors, caused by factors such as noise in the pulse amplitude and duration, a non-harmonic trap and evolution between the `instantaneous' pairs of pulses.  The most significant factor, however, is the experimental limitation on the repetition rates of the necessary ultra-fast lasers.

As $n$ is scaled up, we require increasing numbers of pulses to be incident on the ions in a timescale much smaller than the trap period.  A typical maximum repetition rate for lasers useful for QIP is given by $300$MHz \cite{Kiel11IQEC}.  This restricts the number of pulses where the small timescale approximation holds and the error for the scheme is less than $10^{-4}$, a significant threshold for error correction \cite{NC00, Benh08NP}.  At this repetition rate, the GZC scheme as presented in \cite{GZC03a} does not achieve the requisite fidelity even for $n=1$, while an alternative fast gate scheme \cite{Duan04PRLa} is limited to gate times on the order of the trap period $284$ns \cite{Leib03Nat}.  A faster repetition rate of 1GHz permits the GZC scheme for $n=1$ at twice the trap period, however the $n=2$ case still has too large an error from the assumed instantaneous pulses.

A further restriction is imposed on the timing of ultra-fast laser pulses. The laser emits pulses at regular time intervals, while the GZC scheme requires the pulse group times $t_k$ to be free.  In the next section we will present a pulse splitting technique, making the scheme compatible with the timing and repetition rate restrictions.

\section{Splitting scheme} \label{SSplitting}

Recently, Senko \emph{et al.}~\cite{SeM12} used pulse trains, where large pulses are split into smaller components, to apply a fast spin-dependent momentum transfer to an atom.  This motional control is performed in the strong excitation regime, where the Rabi frequency $\Omega$ is greater than the trap frequency $\nu$.  The pulse splitting turns each laser-emitted pulse into multiple ion-incident pulses using beam-splitters, overcoming the repetition rate restriction.  The technique also gives some pulse timing freedoms; the path lengths for the different pulse components after splitting can be varied.  This technique can be used for significant gains in implementable gate times.

There is an immediate limit on the pulse splitting technique: we are restricted by the maximum pulse area (largest Bloch sphere rotation) that we can generate with our laser.  If each laser pulse can contain energy sufficient for up to 1024 $\pi$ pulses, using the splitting technique this corresponds to a maximum of 512 pairs of $\pi$ pulses.  This is still a significant advantage: the effective repetition rate (incident on the ions) is 512 times the repetition rate of two counter-propagating lasers without pulse splitting.


\subsection{Application to GZC scheme}

\begin{figure}[h!]
     \centerline{\includegraphics[width=0.7\columnwidth]{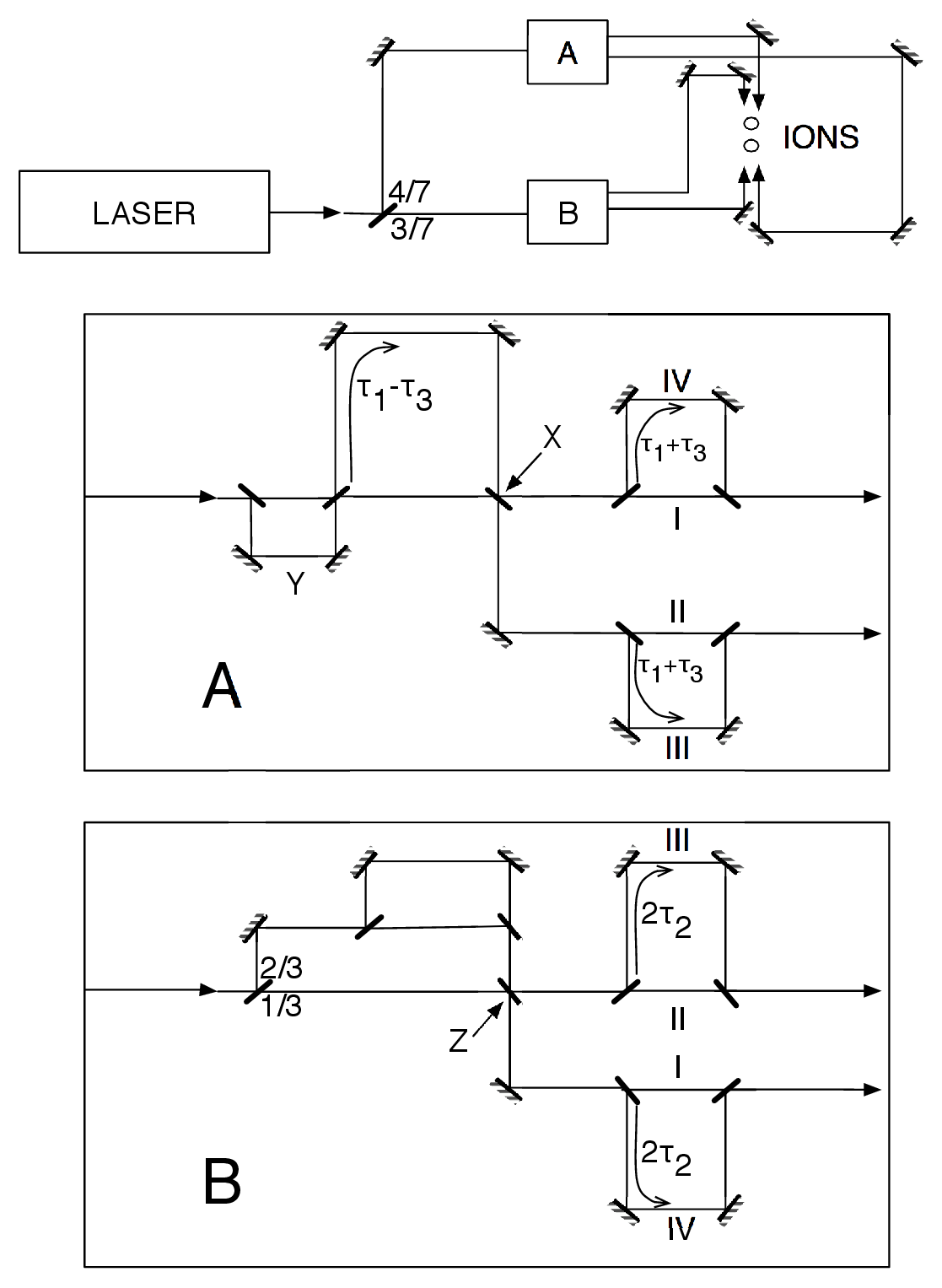}}
   \caption{Optical setup for the GZC gate, with $n=1$.  The pulse delay times, permitted by tailoring path lengths, are marked on the delay loops such that the GZC scheme is performed.  Loops without explicit delays, such as $Y$, are assumed to have zero time delay.  The Roman numerals on particular pulse paths corresponds to the ordering of the pulses.  In the path through $A$, each pulse is split in two at $X$.  Both of the component pulses are subsequently split in two, with delays such that pulse components I and II arrive as a pair, then III and IV (with I arriving immediately before II, and III before IV).  The path through $B$ involves a complex setup before point Z where the delays are as short as possible such that a pulse is split into thirds: three component pulses of equal area reach point Z.  These pulses are split into ordered pairs as with the path through A.  The beam-splitters are 50/50 unless otherwise marked with fractions.  The different paths permit the pulse group numbers, timings and directions as specified by the GZC scheme.}
   \label{GZCsetup}
\end{figure}

The GZC gates become experimentally viable (although complex) using the pulse splitting technique, since both the repetition rate and timing restrictions are relaxed.  We present a possible optical setup for the GZC scheme in Figure \ref{GZCsetup}.  With pulse splitting, the quasi-instantaneous pulse groups can be produced by time delays much smaller than the repetition period using beam-splitter loops, as shown in the figure.  The number of pulses in the GZC scheme scale with $n$.  The figure represents $n=1$, while $n=2^j$ can be implemented by introducing extra beam-splitter loops identical to $Y$, connected to the front of the optical setup (directly before the (4/7):(3/7) beam-splitter) as $Y$ is to the $\tau_1-\tau_3$ delay loop.  We will assume that the delay time for such a loop is zero so that the pulse groups arrive simultaneously. In practice there will be a small delay producing negligible error (details in section \ref{Gerror}).  For loops labelled with delays, such as $\tau_1 - \tau_3$, the marked delay corresponds to the total delay as compared to the straight optical path.

This complex setup would be experimentally challenging to implement, since it was not designed with the pulse splitting technique in mind.  It is possible to use alternate but similarly complicated beam-splitter and mirror arrangements to perform the scheme.  In the following section we remove the GZC symmetries and consider schemes designed for experimental simplicity with the pulse splitting technique, optimising the pulse timing and direction freedoms.

\section{Optimal schemes} \label{SOptim}

\subsection{Cost function}
\begin{figure}[h!]
     \centerline{\includegraphics[width=\columnwidth]{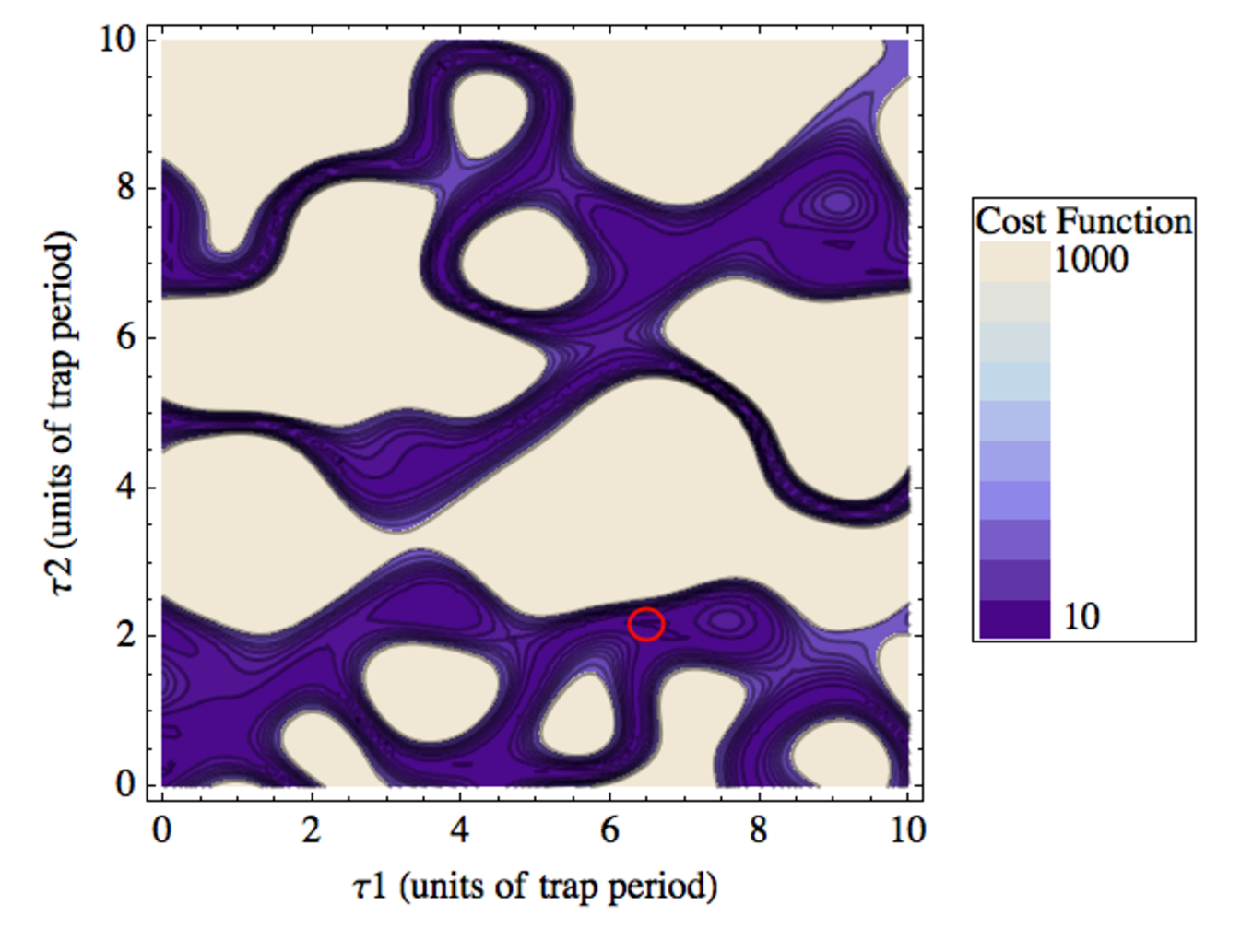}}
   \caption{Solution space shape for $\log(J)$ for our cost function $J$, given the GZC scheme with $n=1$.  The $\tau_3$ value is set to $1.8 T_P$ (trap periods), its solution value.  The global minimum has been circled.}
   \label{convexity}
\end{figure}

To find optimal solutions to the condition equations, we must first introduce an appropriate cost function.  Our cost function $J$ rewards fast gate times $T_G$, balanced with high fidelity, as quantified in section \ref{secPt}.  We write our cost function in terms of the error $E=1-F_P$, where $F_P$ is the process fidelity \cite{Gilchrist2005}:
\begin{eqnarray}
J = \int^{T_G}_0 dt + A \exp [B E], \label{eqcf}
\end{eqnarray}
where $A$ and $B$ are constants introduced to provide a balance between the gate time and fidelity in the cost function.  Simulations using various values of $A$ and $B$ indicated that appropriate values are $A=10$, $B=100$.  The solution space for this cost function is not convex, as shown in Figure \ref{convexity}, thus finding the global minimum is a hard problem.  The figure shows the $\log$ of the above cost function for clearer contours, and the true cost function has the same features.  A useful function involving the condition equations with a convex solution space could not be determined.  The figure shows a 2D-slice of the solution space for the GZC scheme with $n=1$, where $\tau_3$ is set to its solution value associated with the global minimum.  The third dimension further obscures the global minimum, as do extra dimensions for more free variables in later explored schemes, thus searching for solutions (optimal gate schemes) becomes increasingly computationally difficult.

Using our cost function, $J$, we optimised the gate times for different experimental setups using the NLopt nonlinear-optimization package \cite{nlopt}, implementing a controlled random search with local mutation, as outlined in \cite{Kael06JOTA, Pric83JOTA}.

\subsection{Simple scheme: direct pulses}

\begin{figure}[h!]
     \centerline{\includegraphics[width=\columnwidth]{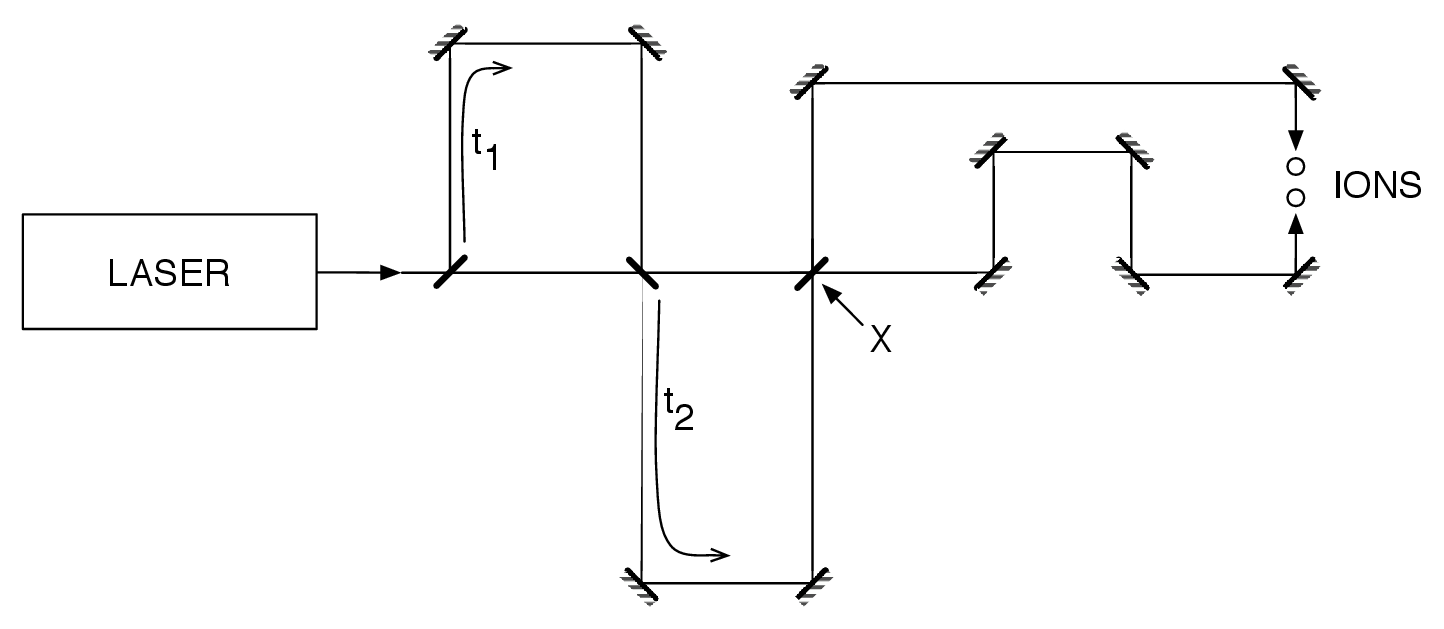}}
   \caption{Optical setup for splitting each pulse into 8 pulses, in 4 pairs, with variable delays $t_1$ and $t_2$.  Each pulse is split into a pair of pulses at $X$, after which one component pulse is incident on the ions from above in the figure, and its counterpart is incident from below.  The path lengths are tailored such that the second pulse in each pair is incident on the ions directly after the first.}
   \label{opticssetup}
\end{figure}

We consider first the simplest setup for the pulse splitting technique.  Ideally, we want to use a single laser to avoid mismatches in pulse timing and frequency, introducing new sources of error.  Since we require pairs of counter-propagating pulses, the simplest setup is in the form of Figure \ref{opticssetup}.  The pulse pairs are formed at the last beam-splitter, $X$.  One of the paths after this beam-splitter is slightly shorter than the other, such that the pulses are incident separately on the ions (with minimal separation in time for each pair).  The first pulse in each pair always arrives from the same direction using this scheme, while in the following section we consider alternating the direction of the first pulse in each pair.  We require resonant $\pi$ pulses to satisfy the simplified controlled phase gate conditions in equation~(\ref{eq:cond}).

Figure \ref{opticssetup} illustrates the pulse splitting process for each laser pulse split into eight components, or eight splittings, which corresponds to four counter-propagating pulse pairs.  More splittings are easily produced: the number of splittings is doubled when an extra beam-splitter loop, of the form shown with a $t_1$ delay in the figure, is introduced.  Each added loop links to the others in the same way the $t_1$ delay loop is joined to the $t_2$ delay loop.  These delay times are optimised to find the fastest gate schemes.

The optimal gate schemes for this setup provide gate times on the order of the trap period; the fastest gate found is 1.37$T_P$ (389ns), where $T_P$ is the trap period.  This gate is for one pulse split into eight pairs, and the next fastest gates are around twice the trap period.  We use a Lamb-Dicke parameter of $\eta = 0.2$ \cite{Leib03Nat}, and a repetition rate of 300MHz.

It is significant to note that each solution set of optimised delay times for different numbers of pulses and splittings contained an exact multiple of $0.5$ and of $\frac{1}{2\sqrt{3}}$.  This is equivalent to a kick with direction $(-1)$ in the rotating centre of mass and stretch mode phase space trajectories respectively.  It follows that the positive direction of each pulse pair kick results in closed phase space trajectories only when the evolution time permits an effective kick reversal.  This explains the restriction of the scheme timescale to the order of the trap period.  We thus consider a scheme with little extra complexity that permits the direction reversal of the first pulse in a pair.

\subsection{Simple scheme: alternating pulses} \label{sap}

\begin{figure}[h!]
     \centerline{\includegraphics[width=\columnwidth]{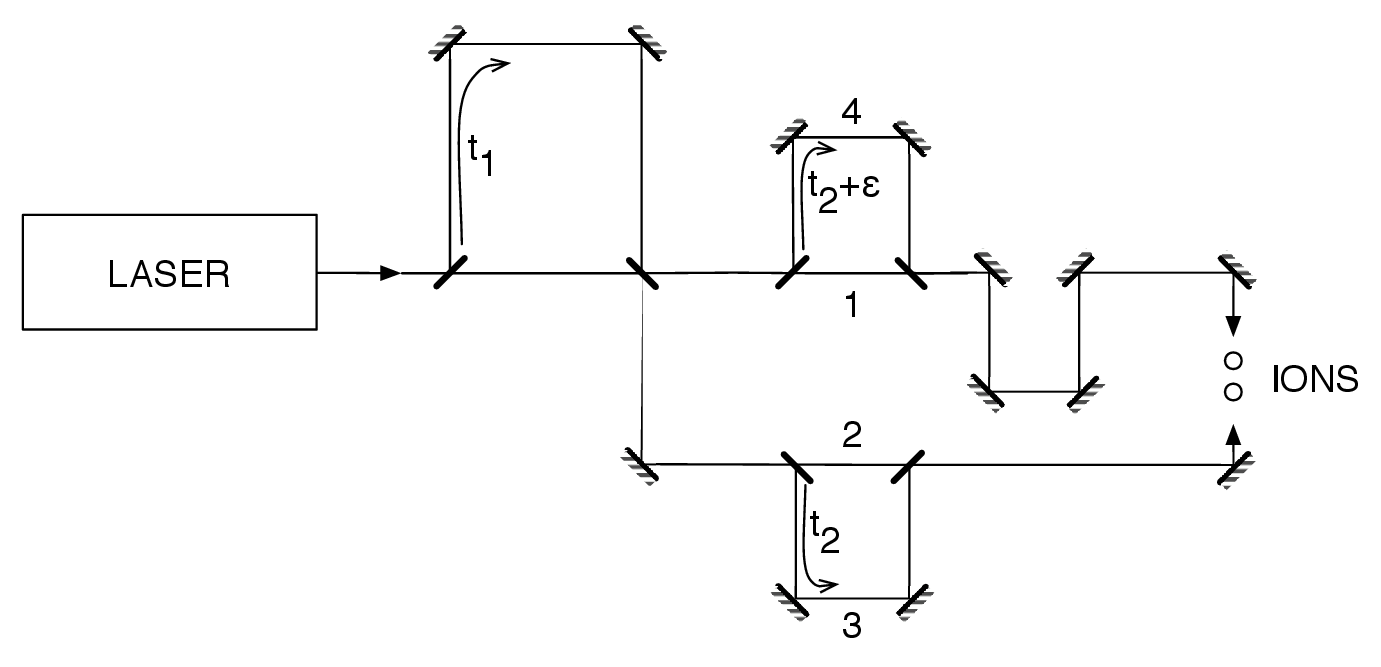}}
   \caption{Optical setup for splitting each pulse into four pairs of pulses, with variable delays $t_1$ and $t_2$.  The numbers correspond to the arrival order for the pulses, as for the GZC scheme.  Note that the first pulse of the first pulse pair arrives from the top, while the first pulse of the second pulse pair, number 3, arrives from the bottom.  Thus the pulse pairs have two possible direction orderings.}
   \label{opticssetupb}
\end{figure}

\begin{figure}[h!]
     \centerline{\includegraphics[width=\columnwidth]{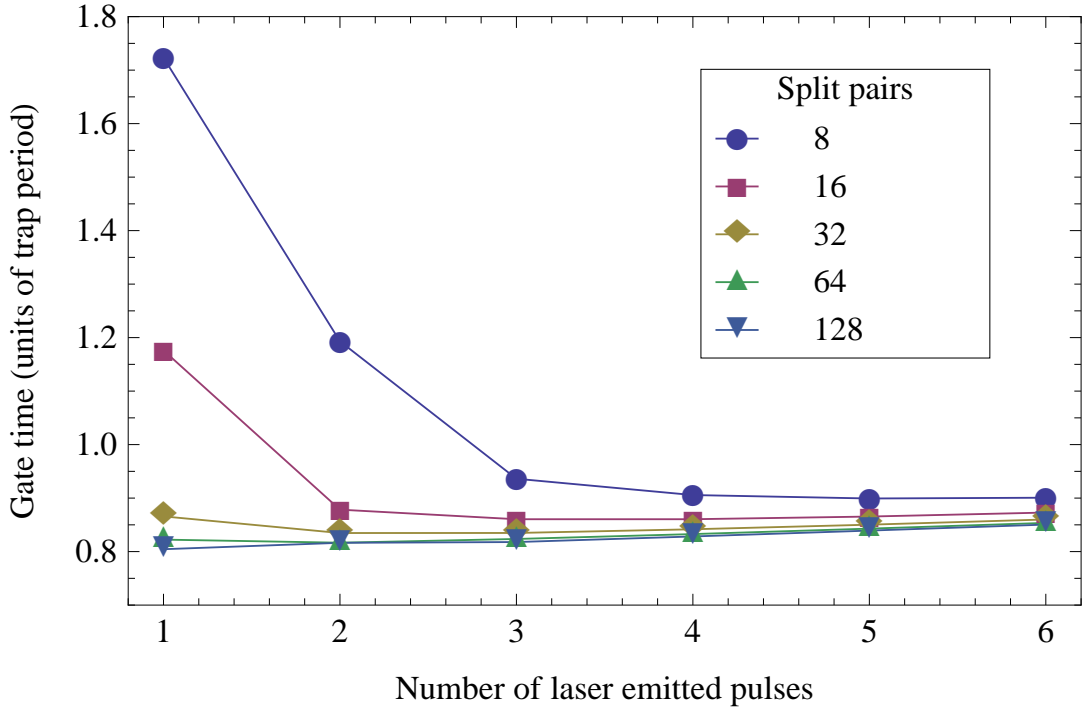}} 
   \caption{Gate operation time for different numbers of pulses, using the alternating pulse direction scheme. Each curve corresponds to a different number of splittings of each laser pulse, corresponding to the indicated number of split pairs.  The curves approach a line with a slight gradient, which represents the limiting gate time for the scheme.  The slope of the line is governed by the laser repetition rate, 300MHz.}
   \label{AltScaling}
\end{figure}

When a direction switch is introduced for certain pairs of pulses, as in Figure \ref{opticssetupb}, there is a small speedup in the gate time to below the trap period.  Two splitting loops with slightly different lengths ($t_2$ and $t_2+\epsilon$ in the figure) allow the ordering of the counter-propagating pulses to be reversed.  Figure \ref{AltScaling} shows the gate times for different numbers of pulses.  Different numbers of splittings of each emitted pulse are compared, where each number of splittings corresponds to an experimental setup, with more splittings introducing extra beam-splitter loops.

Note that each setup approaches a gate time lower limit around $0.8T_P$.  Larger numbers of splittings, corresponding to more power applied to the ions, approach this limit faster.  As for the direct pulse scheme, we are restricted by the trap period timescale, since each solution involves delay times $0.5 + \frac{1}{2 \sqrt{3}} = 0.79T_P$.  The lower limit increases slowly with the number of pulses, as the laser repetition rate is taken into account.  In principle, the fastest gate corresponds to sending one pulse with very large area, which is split into a large number of pulses performing a phase gate.  In practice, the number of splittings is restricted by the laser capabilities.  For a given maximum pulse energy, the number of possible splittings is fixed and the fastest gate is achieved by some optimal number of pulses, determined by the minimum for each curve in Figure \ref{AltScaling}.  Our optimisation considered up to 128 pairs of pulses, which requires 11 beam-splitters.


The limiting timescale indicates that introducing a reverse direction pulse pair on a single delay loop doesn't provide the required freedom for schemes with scalable gate times as the applied power is increased.  In phase space, this represents the requirement for this alternating scheme to have trap period delay times to close the phase space trajectories.  The GZC scheme is not limited by the trap timescale, which means that appropriate placement of the reverse pulse pairs removes the timescale limitation.  We identify optimal reverse pulse pair timings through a free search of solution space, and reverse engineer the practical experimental setups from these solutions.


\subsection{Reverse Engineering}

The three condition equations suggest that three variables are required for solutions, however added free variables can introduce further flexibility and reduce the cost function for more optimal schemes.  It was found for the simple setup schemes in the previous two sections that even when extra delay loops, and corresponding variables, were added, only three of the delay times took nonzero values for optimal solutions.  This suggests that the symmetry of the scheme plays a significant role.  Intuitive design of simple optical setups with gate times that scale with power proved challenging.  We instead use a free search of solution space and identify symmetries leading to implementable setups.  The limitations of global optimisation searches restrict our search to low numbers of free variables.

Free variable searches were carried out for optimised times $\underbar{t} = (0, x_1, x_2, x_3, x_4, x_5, ...)$ associated with kicks $\underbar{z} = (1, -1, 1, -1, 1, -1, ...)$.  A search over five free variables revealed that the optimal solution takes the following symmetric form with $(a,b,c)=(1,1,1)$:  
\begin{subeqnarray} \label{eqsymmf}
\underbar{t} = (-\tau_1, -\tau_2, -\tau_3, \tau_3, \tau_2, \tau_1) \\
\underbar{z} = (an, -bn, cn, -cn, bn, -an).
\end{subeqnarray}
As before, $|z_k|$ is the number of pairs of pulses applied to the ions with first pulse direction $\sgn(z_k)$ at time $t_k$.  This $(1,1,1)$ scheme, though straightforward to implement, is still limited by the trap period, approaching a gate time around $0.73T_P$.

Nine free variables give more variety in optimal solutions.  For 8 split pairs, a symmetric scheme in the form of equation (\ref{eqsymmf}) with $(a,b,c)=(1,2,2)$ was found to be optimal.  Higher numbers of split pairs gave non-symmetric optimal solutions, such as a scheme for 32 split pairs (320 total incident pulse pairs) with a gate time of $0.086T_P$.  

The non-symmetric schemes are extremely complex to implement using the beam-splitter loops.  It also becomes hard to find the equivalent schemes for higher numbers of splittings, since the global optimisation search is not deterministic, and we have a complicated high-dimensional search space.  In contrast, the general symmetric form in equation (\ref{eqsymmf}) provides a lower-dimensional search space, and certain values of $(a,b,c)$ for the symmetric form permit straightforward experimental setups.  Note that this symmetric form includes the GZC solution, which takes the values $(a,b,c)=(2,3,2)$.  

\begin{figure}[h!]
     \centerline{\includegraphics[width=0.7\columnwidth]{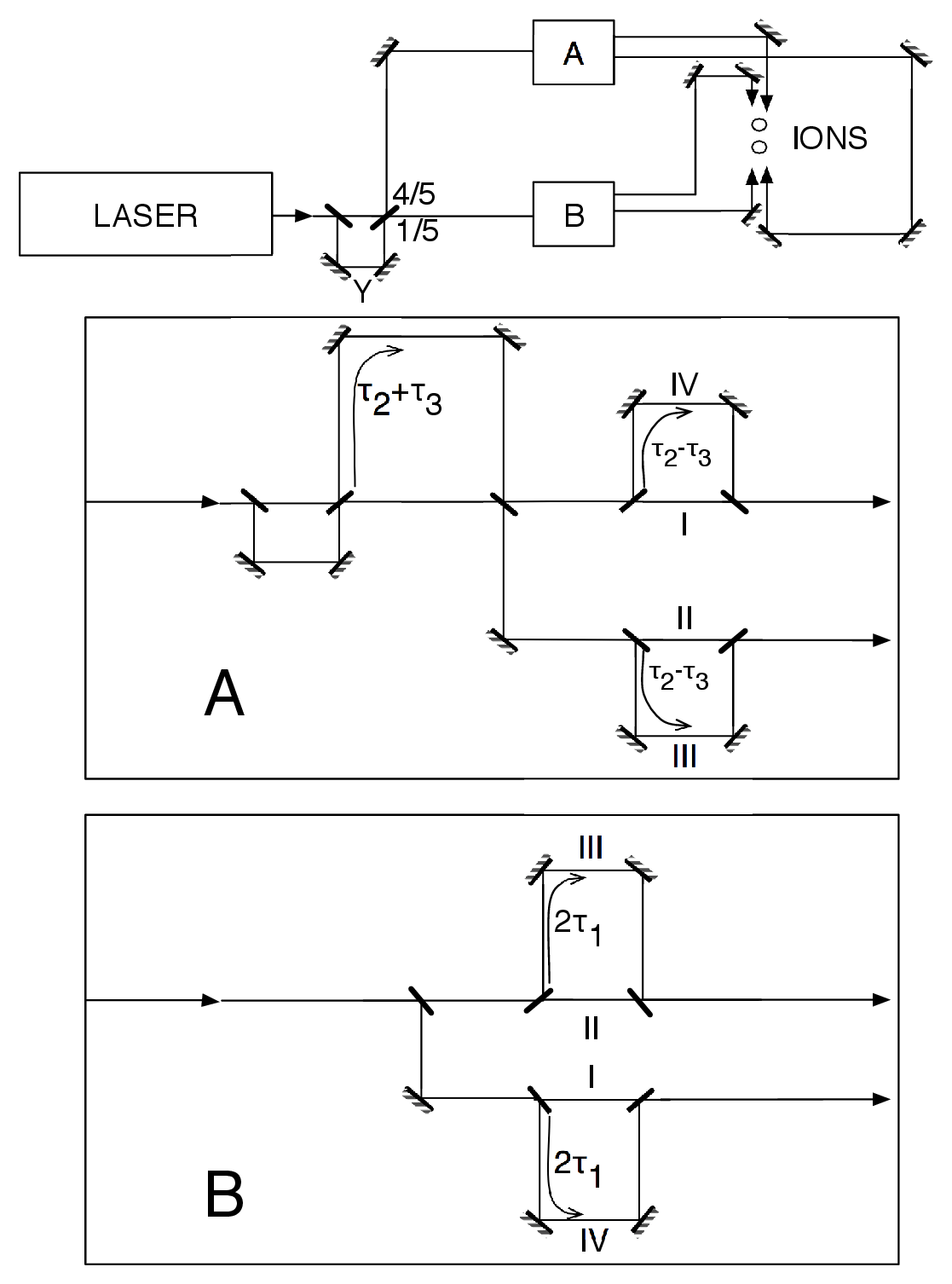}}
   \caption{Optical setup for the $(a,b,c)=(1,2,2)$ scheme for $n=2$, \emph{i.e.} 20 incident pulse pairs.  The roman numerals mark the order of incidence for pulses travelling through the marked path.  Beam-splitter loops without a marked time delay represent a negligible delay.  The loop marked $Y$ doubles the number of splittings; without it we would have the $n=1$ case, while another such loop would give us $n=4$.}
   \label{122setup}
\end{figure}

These symmetric schemes are often well suited to solving the condition equations such that the gate time is not limited by the trap period, as for the GZC case.  The schemes, although sometimes suboptimal, are among the better solutions for the free variables searched.  Searching low integer values for $(a,b,c)$ to find implementable, scalable schemes yielded the $(1,2,2)$ scheme as the optimal experimentally implementable scheme.  Figure \ref{122setup} shows the optical setup of the $(1,2,2)$ scheme.  It is similar in structure to the GZC setup, with some complexity removed.

\begin{figure}[h!]
     \centerline{\includegraphics[width=1.0\columnwidth]{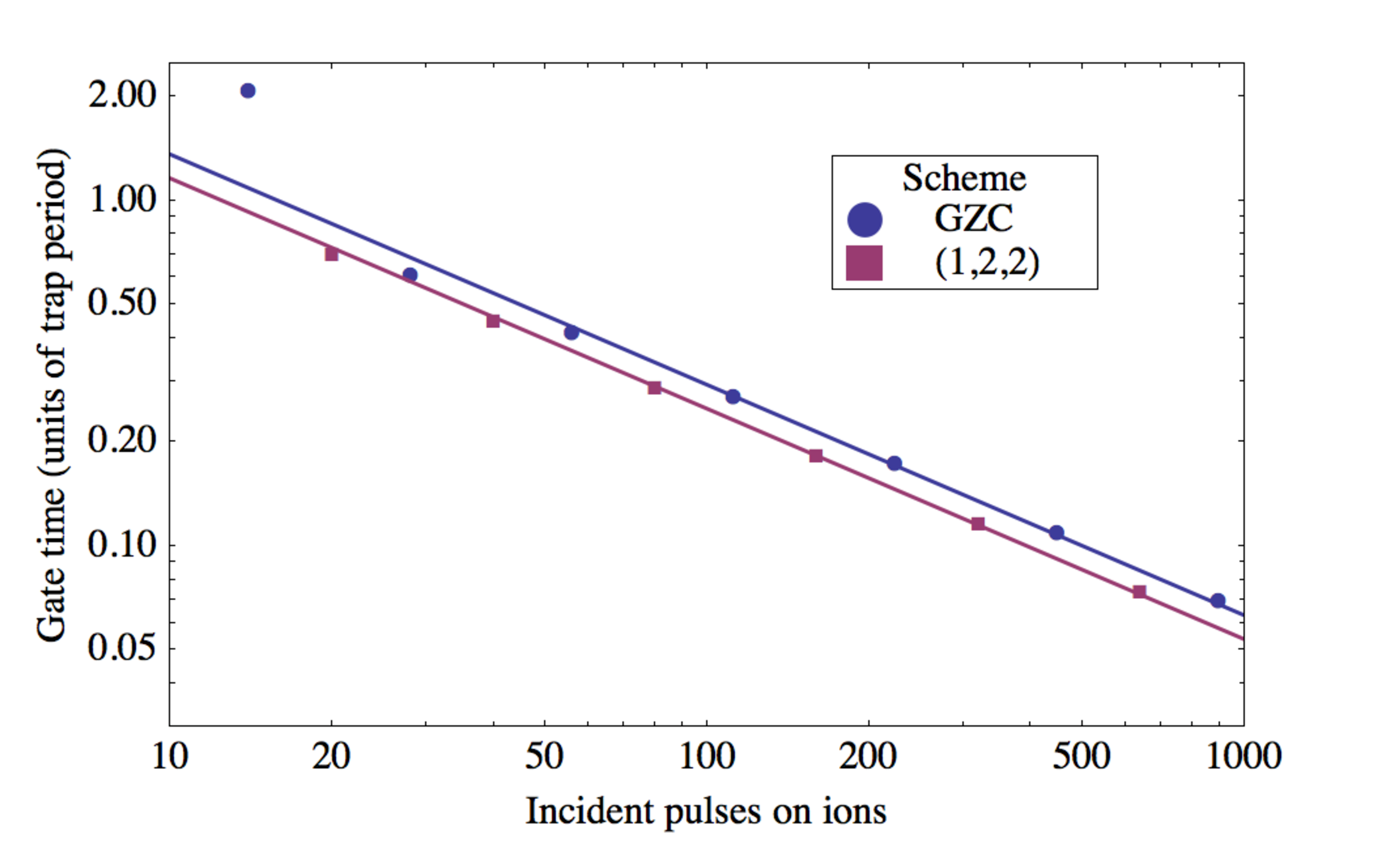}}
   \caption{Gate times for the $(1,2,2)$ symmetric scheme and the GZC scheme for different numbers of pulses incident on the ions.  Both schemes are performed by a single laser emitted pulse split into the required incident pulse pair components.  The linear fits represent the power law scaling for each scheme, gate time $T_G  \propto N^{-2/3}$ for $N$ pulse pairs.  The scaling coefficients are 6.30 and 5.37 for the GZC and $(1,2,2)$ schemes respectively.}
   \label{compareplot}
\end{figure}

As derived in \cite{GZC05a}, the optimal relationship between applied pulse pairs $N$ and gate time $T_G$ follows the power law $T_G  \propto N^{-2/3}$.  The $(1,2,2)$ scheme and the GZC scheme both follow this scaling for large numbers of pulses, with a slightly better prefactor (15\% smaller) for the $(1,2,2)$ scheme as shown in Figure \ref{compareplot}.  Lower numbers of splittings approach this scaling at different rates for either scheme, seen in the figure.

The scaling suggests infinitesimal gate times as the number of incident pulse pairs increases towards infinity.  The setup for each scheme involves a single pulse split into the required components for the entire gate operation, which means the number of incident pulse pairs are directly linked to the laser emitted pulse energy.  The prior limitations in fidelity due to finite laser repetition rates are thus transferred to a gate time limitation due to finite laser powers.  This power limitation leaves enormous scope for experiments, however.  The $(1,2,2)$ scheme requires an initial pulse delivering $2 \sqrt{10} \pi$ for 10 incident $\pi$ pulse pairs ($n=1$).  An achievable emitted picosecond pulse (such that the pulse duration is negligible relative to the trap period) could deliver around $100 \pi$, which would correspond to 2500 incident pulse pairs ($n \simeq 256$).  Stronger limitations on gate times arise from fidelity considerations, which we discuss in section \ref{secPa}.

We have considered multiple schemes with different setup complexities and scaling relationships.  The fastest scheme for less than 20 incident pulse pairs is the simple alternating pulse scheme (section \ref{sap}), which provides a gate time of $1.18T_P$ for 16 incident pairs, given a very simple setup.  For 20 or more incident pulse pairs, some extra complexity gives gate times scaling most efficiently with the laser power for the $(1,2,2)$ scheme.  This scheme achieves a gate time of $0.12T_P$ (33ns) for 320 incident pulse pairs.  This is more than two orders of magnitude faster than experimental gates with ion trap systems \cite{DeMa02PRL, Kirch09NJP}.

\section{Limitations} \label{Gerror}

Gate fidelity loss is caused by imperfect laser control, as well as dissipation from coupling to the environment.  The effects of environmental decoherence are considered in \cite{GZC05a}, and were found to cause exponential decay of the fidelity.  However, the dissipation limit on gate speeds was found to not restrict our scheme.  We find that control errors introduce a stronger bound on gate speed.  Imperfect laser control encompasses the effects of finite pulse durations, non-instantaneous evolution between pulses, mistimed pulses, and errors in the amplitude and duration of each pulse.  

\subsection{Pulse timing} \label{secPt}

Using the condition equations (\ref{Phase},\ref{Cc},\ref{Cr}), we determine the significance of shifting pulses in time, analysing the effect of mistimed pulses and pulses assumed to be instantaneous (using zero as a beam-splitter loop delay time).  The error from failing to satisfy the motional mode conditions is derived in \cite{GZC05a}:
\begin{eqnarray}
C_1 \simeq \exp [ -\frac{1}{2} |2 \eta_c C_c |^2 ] \\
C_2 \simeq \exp [ -\frac{1}{2} |\eta_r C_r |^2 ] \\
E_m = (6 - C_1^4 - C_2^4 - 4 C_1 C_2 )/8.
\end{eqnarray}
Here we have assumed a pure initial internal state, separable from an initial low energy thermal motional state.  Ideally, the internal and motional states become entangled during the gate, before the components are again separable upon the gate's completion.  The error is found using the evolution of the internal state density matrix coefficients, detailed in \cite{GZC05a}.

The upper bound phase error for the process fidelity when the acquired phase $\Theta = \pi/4 - x$, for some error $|x| < \pi/4$, is given by
\begin{eqnarray}
E_p \leq \frac{3}{4} - \frac{3}{4} \cos(2x).
\end{eqnarray}

The error $E=E_m+E_p$ given here is defined as $1-F_P$, where $F_P$ is the process fidelity.  The total error for failing to satisfy the conditions in Eqs.~(\ref{eq:cond}) is thus given by
\begin{eqnarray}
E \leq \left(\frac{3}{4} - \frac{3}{4}\cos(2x)\right) + \frac{1}{8}\left(6 - C_1^4 - C_2^4 - 4 C_1 C_2\right). \label{conderror}
\end{eqnarray}

The pulse timings were systematically shifted, and it was found that for an error upper bound of $10^{-4}$, the system was stable for systematic time shifts of beyond 14ps (using a repetition rate of 300MHz, trap frequency of $2 \pi \times 3.52$MHz).  Pulse duration can be much smaller than this value, and this corresponds to a systematic mirror placement error of 0.4cm, which is much larger than the accuracy achievable in experiments.  The pulse time shift errors are thus not a limiting factor for this scheme.

\subsection{Pulse area} \label{secPa}

Pulse duration and amplitude errors, on the other hand, can be significant.  Both duration and amplitude errors affect the area of the pulse sent, which affects the area of each pulse component in the splitting scheme.  The incident pulses on the ions will not be exact $\pi$ pulses, meaning that the condition equations are no longer applicable.  We consider the worst case process fidelity $F_W$ \cite{Gilchrist2005}:
\begin{eqnarray}
F_W = \min_\psi (| \left< \psi \right| U_I^\dagger U_\epsilon \left| \psi \right> |^2), \label{eqwfid}
\end{eqnarray}
where $\psi$ is the quantum state of the ions, thus we are minimising over motional and internal states.  $U_\epsilon$ is the realised unitary operator for the process with errors, and $U_I$ is the ideal gate scheme unitary operator,
\begin{eqnarray}
U_I= e^{i \frac{\pi}{4} \sigma_1^z \sigma_2^z}.
\end{eqnarray}

Recall that this ideal unitary is composed of momentum kicks for each ideal pair of pulses ($z_1 = \pm1$):
\begin{eqnarray} 
U_{Ip}=e^{-2ikz_1 (x_1\sigma_1^z + x_2\sigma_2^z)},
\end{eqnarray}
interspersed by the free evolution of the system, which rotates the motional modes in phase space:
\begin{eqnarray}
U_r = e^{-i \nu \delta t_k (a^\dagger_c a_c + a^\dagger_r a_r)}.
\end{eqnarray}

For a general atom-laser interaction, if $\theta$ is the rotation about the Bloch sphere caused by the pulse of duration $T$, i.e. $\theta = \int_0^{T} \frac{\Omega (t)}{2} dt$, the unitary is given by
\begin{eqnarray}
	U_{\mathrm{kick}} &= (\cos(\theta) - i\sin(\theta)\sigma_1^x e^{-ikx_1\sigma_1^z}) (\cos(\theta) - i\sin(\theta)\sigma_2^x e^{-ikx_2\sigma_2^z}) \label{equkick}
\end{eqnarray}

To examine $\pi$ pulses with some small error, we consider $\theta = \frac{\pi}{2}+\epsilon$, $\epsilon \ll \frac{\pi}{2}$.  The unitary $U_p$ for a pair of counter-propagating pulses can be expanded in terms of $\epsilon$:
\begin{eqnarray}
U_p = U_{Ip} + \sum_j \epsilon^j f_j, \label{equexp}
\end{eqnarray}
where each $f_j$ is found from equation (\ref{equkick}).  The unitary for a pulse scheme is given by combinations of free evolution, $U_r$, and pairs of pulses, $U_p$.  For a four pulse pair scheme:
\begin{eqnarray}
U_\epsilon = U_p U_r U_p U_r U_p U_r U_p. \label{equeps}
\end{eqnarray}

We considered the worst case fidelity to second order in $\epsilon$, as the first order terms cancel.  For the four pulse pair scheme, a lower bound on the worst case fidelity was found to be $F_W = 1- 331\epsilon^2$.  This means that for an error of $1\%$ in the $\pi/2$ pulse energy, for example, $\epsilon=5\times 10^{-3}$ and the worst case fidelity is $99.2\%$.



The fidelity can scale very poorly as the number of pulses increases, as expected for systematic errors in the pulse area.  Substituting equation (\ref{equexp}) into equation (\ref{equeps}) gives four terms ($^4_1 C$) to first order in $\epsilon$:
\begin{eqnarray}
\eqalign{U_\epsilon &= U_I + \epsilon (f_1 (U_r U_{Ip})^3 + U_{Ip} U_r f_1 (U_r U_{Ip})^2 + (U_{Ip} U_r)^2 f_1 U_r U_{Ip} \\&+ (U_{Ip} U_r)^3 f_1) + \mathcal{O}(\epsilon^2).}
\end{eqnarray}
This corresponds to permutations of $f_1$ and $U_{Ip}$.  Thus for an $N$ pulse scheme there are $^N_1 C = N$ such terms.  Similarly there are $N$ second order terms of $U_\epsilon$ corresponding to permutations of $f_2$, and $^N_2 C$ second order cross terms in $f_1$.  Higher orders $m$ of $\epsilon$ will contribute $^N_m C$ cross terms in $f_1$.  For large $N$, this scales as $N^m$.   Note that for $\epsilon$ on the order of $0.01$, and $N$ on the order of 100, perturbation theory breaks down, indicating that the gate errors become large.

Pulse area errors are thus a limiting factor for high fidelity fast gates.  We have considered an absolute worst case bound, and experimental error is expected to be much less than this bound.  The scaling of error with pulse numbers is nonetheless concerning.  These errors are not unique to pulse splitting schemes; the pulse area errors must be considered for any gate scheme requiring a given pulse area and multiple pulses.  Techniques such as stimulated Raman adiabatic passage (STIRAP) \cite{GZC05a, Berg98RMP} exist for limiting pulse control errors, which would greatly improve the fidelity of gate schemes, although possibly at the cost of gate speed.  Even with low pulse numbers, we can achieve gate times faster than the trap period.

\subsection{Pulse direction} 

The angular stability of the lasers also plays a part in the fidelity of the scheme.  Systematic errors were introduced independently for the two pulse directions to represent misalignment with the axis of the ion trap.  As one would expect, the sensitivity to pulse direction becomes more significant with the number of pulses.  The $(1,2,2)$ scheme for 20 pulses has error less than $10^{-4}$ for angular precision of $0.36^\circ$, or 6.3 milliradians.  The scheme with 80 pulses requires stability to $0.09^\circ$, or 1.6 milliradians.  Again we see that errors are compounded from increasing the number of pulses.

\section{Conclusions}

We have considered gate schemes optimised for both gate time and experimental simplicity using the pulse splitting technique.  The most practical optical setup schemes are limited by the trap period timescale, and achieve gate times of up to $0.8T_P$ (0.23$\mu$s).  Some added complexity provides gate times that scale with the number of pulses applied to the ions.  Our $(1,2,2)$ scheme is limited by control errors such as laser intensity and duration fluctuations, however gate speeds faster than the trap period can be achieved with high fidelity.  For 80 incident pulse pairs, the scheme achieves gate times of 0.29$T_P$ (82ns), more than two orders of magnitude faster than experiments, and more pulses give faster gate times.  For comparison, an alternate fast gate scheme \cite{Duan04PRLa}, using directly incident sequences of $\pi$ pulses, has an operation time longer than 100ns for the same 300MHz repetition rate.

The techniques explored for the gate implementation can be extended to consider the excitations of higher numbers of ions by introducing new control parameters to account for the added motional modes.  Alternatively, the gate could be implemented in a shuttling architecture \cite{Kiel02Nat}.  When thus extended, the gate scheme presented would be implementable for more complex quantum information processing operations.

\ack
ARRC thanks the support from the Australian Research Council Centre of Excellence for Quantum Computation and Communication Technology (Project number CE110001027).  DK was supported by an Australian Research Council Future Fellowship (FT110100513).  JJH was also supported by an Australian Research Council Future Fellowship (FT120100291).  The work was also supported by DP130101613 (DK, JJH, ARRC).

\section*{References}
\bibliographystyle{iopart-num} 
\bibliography{library2}

\end{document}